\documentclass[aps,prl,onecolumn,english,balance,superscriptaddress,floats,showpacs,letter,twocolumn]{revtex4-2}
\usepackage[T1]{fontenc}
\usepackage[latin9]{inputenc}
\usepackage{amsmath}
\usepackage{amssymb}
\usepackage{stmaryrd}
\usepackage{graphicx}
\usepackage{esint}
\usepackage{subfigure}
\usepackage{multirow}
\usepackage{xcolor}
\usepackage{ulem}
\usepackage{accents}

\makeatletter

\newcommand{\beq}{\begin{equation}}
	\newcommand{\eeq}{\end{equation}}
\newcommand{\bea}{\begin{eqnarray}}
	\newcommand{\eea}{\end{eqnarray}}
\newcommand{\bwt}{\begin{widetext}}
	\newcommand{\ewt}{\end{widetext}}
\@ifundefined{textcolor}{}
{%
	\definecolor{BLACK}{gray}{0}
	\definecolor{WHITE}{gray}{1}
	\definecolor{RED}{rgb}{1,0,0}
	\definecolor{GREEN}{rgb}{0,1,0}
	\definecolor{BLUE}{rgb}{0,0,1}
	\definecolor{CYAN}{cmyk}{1,0,0,0}
	\definecolor{MAGENTA}{cmyk}{0,1,0,0}
	\definecolor{YELLOW}{cmyk}{0,0,1,0}
}

\newcommand{\bk}{\mathbf{k}}

\newcommand{\bA}{\mathbf{A}}

\newcommand{\bp}{\mathbf{p}}

\newcommand{\br}{\mathbf{r}}

\newcommand{\bx}{\mathbf{x}}
\newcommand{\by}{\mathbf{y}}
\newcommand{\bz}{\mathbf{z}}

\newcommand{\eps}{\epsilon}

\newcommand{\mbf}[1]{\mathbf{#1}}

\newcommand{\mcH}{\mathcal{H}}
\newcommand{\mcC}{\mathcal{C}}

\newcommand{\fvec}[1]{\boldsymbol{#1}}

\newcommand{\half}{\frac{1}{2}}
\newcommand{\pd}{\partial}
\newcommand{\rmd}{{\rm d}}

\makeatletter
\newcommand{\doublewidetilde}[1]{{%
  \mathpalette\double@widetilde{#1}%
}}
\newcommand{\double@widetilde}[2]{%
  \sbox\z@{$\m@th#1\widetilde{#2}$}%
  \ht\z@=.9\ht\z@
  \widetilde{\box\z@}%
}
\makeatother

\usepackage{babel}

\makeatother

\usepackage{babel}

\begin{document}

\title{Orbital magnetization and magnetic susceptibility of interacting electrons}

\author{Jian Kang}
\email{kangjian@shanghaitech.edu.cn}
\affiliation{School of Physical Science and Technology, ShanghaiTech University, Shanghai 201210, China}

\author{Minxuan Wang}
\email{minxuan-wang@princeton.edu}
\affiliation{Department of Physics, Princeton University, Princeton, New Jersey 08544, USA}

\author{Oskar Vafek}
\email{vafek@umn.edu}
\affiliation{W.~I.~Fine Theoretical Physics Institute and School of Physics and Astronomy,
University of Minnesota, Minneapolis, Minnesota 55455, USA}

\begin{abstract}
We present a rigorous derivation of the orbital magnetization for interacting electrons within the self-consistent Hartree-Fock approximation. Our method also allows us to derive formulas for the orbital magnetic susceptibility. The results are expressed entirely in terms of the self-consistent wavefunctions and the Hartree-Fock energy spectrum at zero magnetic field. We find that the formula for the orbital magnetization is the same as in the non-interacting case, provided that the Hamiltonian and Bloch wave functions are replaced by their  Hartree-Fock counterparts. By contrast, the orbital magnetic susceptibility contains an additional interaction-induced contribution that cannot be obtained by such a replacement. We test the formulas on an interacting Rashba model, finding an agreement with calculations performed at a small but non-zero external magnetic field.
\end{abstract}
	
\maketitle







Recent experiments on van der Waals heterostructures revealed orbital magnetism~\cite{DavidAH, YoungQAH, Efretov2019, FengWang20, Xiaodong21, MatthewTMBG, KinFaiFCI, JuLongChiralSC, Xiaobo25, FeldmanHofstadter, YoungOrbitalCI, YoungMoTe2}, a class of phenomena that includes (quantum) anomalous Hall effects~\cite{DavidAH,YoungQAH} and hysteretic valley switching~\cite{JuLongChiralSC, YoungOrbitalCI, GuoruiSwitchChern}.
Interestingly, the orbital magnetism,  and the associated spontaneous time reversal symmetry breaking, are induced by electron-electron (e-e) interactions. Owing to their non-zero magnetic moment, the energy of such states changes linearly in a small external magnetic field, $B$. The $B$ field can therefore be used to directly manipulate a given state~\cite{YoungOrbitalCI,JuLongChiralSC, Xiaobo25}. However, due to the prominent role of the e-e interactions, it is challenging to reliably compute the value, or even a sign, of the orbital magnetization in the limit of vanishing $B$.

In the non-interacting case the formula for the orbital magnetization was first derived by solving the semiclassical equations of motion at $B\neq 0$~\cite{xiao2005berry, xiao2010berry} or by employing the transformation between the Wannier and the Bloch basis~\cite{thonhauser2005orbital, ceresoli2006orbital, thonhauser2011theory}. Later, several alternative methods were applied to rederive the formula for the orbital magnetization~\cite{shi2007quantum, PALeeOrbMag, raoux2015orbital, YuxuanBosonization} and to obtain the expression for the orbital magnetic susceptibility~\cite{gao2015geometrical, raoux2015orbital}. This was done by either treating the spatially varying $B$ field perturbatively~\cite{shi2007quantum} or introducing the gauge-invariant Green functions~\cite{PALeeOrbMag,raoux2015orbital}.
However, generalizing the results to interacting systems has proven to be difficult.
When the interactions are treated via the self-consistent Hartree-Fock (HF) approximation at $B\neq 0$, the calculations are significantly more complicated than at $B=0$, because at $B\neq 0$ one must take into account the interactions within magnetic subbands~\cite{wang2023theory, wang2024phase}.  It is computationally challenging to extend this approach to a small $B$ due to the growing number of magnetic subbands. For some orbital magnets, small $B$ regime can extend to several Tesla. For example, this is the case in twisted bilayer MoTe$_2$ at an angle of $3.89^{\circ}$, where, even a $10$Tesla magnetic field threads only $0.057$ of the magnetic flux quantum $\phi_0=hc/e$ per moire unit cell~\cite{wang2024phase}. It is therefore desirable to develop an alternative approach to the interacting orbital magnets that departs from $B\rightarrow 0$ limit and that would be computationally significantly more efficient.

In this work we develop such an approach.
We present formulas for both the orbital magnetization and the orbital susceptibility for the interacting system at zero temperature within the self-consistent HF approximation which involve only the self-consistent solution of the $B=0$ problem. Our main results are presented in the Eqs.~(\ref{eq:M_main_result}) and (\ref{eq:chi_main_result}-\ref{Eqn:ChiInt}). The rest of the paper is devoted to the details of the derivation of the orbital magnetization and the application of the results to a model that allows us to test it explicitly at $B\neq 0$.

We start with a general interacting Hamiltonian
\begin{eqnarray}\label{eq:generalH}
&&\mathcal{H}=\int \rmd\br\ \psi_a^\dagger(\br)\left(\hat{H}_{ab}\left(p_\mu+\frac{e}{c}A_\mu(\br)\right)+U_{ab}(\br)\right)\psi_b(\br)+\mathcal{V},\nonumber\\
&&\mathcal{V}=\frac{1}{2}\int \rmd\br \rmd\br'\ V_{ab,cd}(\br-\br')
\psi_a^\dagger(\br) \psi_c^\dagger(\br')\psi_{d}(\br')\psi_{b}(\br)
\end{eqnarray}
where $\hat{H}+U$ is a single particle Hamiltonian, $\hat{H}$ is a differential operator that acts to the right, which we assume can be expanded as
$\hat{H}_{ab}(p_\mu)=\sum_n
\Lambda_{n,ab}^{\mu_1 \mu_2\ldots\mu_n }p_{\mu_1}
p_{\mu_2}\ldots p_{\mu_n}$,  and where $\Lambda$ is symmetric under the exchange of any two $\mu$ indices (we use the usual summation convention throughout); $U_{ab}=U^*_{ba}$ and the momentum operator $p_\mu=\frac{\hbar}{i}\frac{\partial}{\partial r_\mu}$.
The fermion fields $\psi_a$ satisfy the usual anticommutation relations and may carry internal indices $a$ for say, spin, valley etc. The magnetic field $B$ enters via the static external vector potential $\bA(\br)$ that satisfies $\nabla\times \bA(\br)=B\hat{\bz}$. 
The interaction term is Hermitian therefore $V_{ab,cd}(\br)=V^*_{ba,dc}(\br)$ and
the Fermi statistics gives $V_{ab,cd}(\br)=V_{cd,ab}(-\br)$.
Note that because we are interested in the orbital contribution, the $B$ field couples entirely via the minimal substitution.  

Within the HF method the ground state is approximated by a Slater determinant. The total energy is therefore
\begin{eqnarray}
&&\langle \mathcal{H}\rangle =\nonumber\\ 
&&\int \rmd\br \rmd\br'\left[\delta(\br-\br')\left(\hat{H}_{ab}\left(p_\mu+\frac{e}{c}A_\mu(\br)\right)+U_{ab}(\br)\right)P_{ba}(\br,\br')\right.\nonumber\\
&&\left.+ \frac{1}{2} V_{ab,cd}(\br-\br')
\left(P_{ba}(\br,\br)P_{dc}(\br',\br')-P_{da}(\br',\br)P_{bc}(\br,\br')\right)\right]
\nonumber\\
\end{eqnarray}
where the single particle density operator is defined as the ground state expectation value
$P_{ab}(\br,\br')=\langle \psi_b^\dagger(\br')\psi_a(\br) \rangle$ that can be used to obtain the total particle number $N=\int \rmd \br\ P_{aa}(\br,\br)$. 

The HF Hamiltonian operator is determined by the action of $\frac{\delta \langle \mathcal{H}\rangle}{\delta P_{ba}(\by,\bx)}$  on a wavefunction, giving
\begin{eqnarray}
\hat{\mathcal{H}}^{\text{HF}}_{ab}\psi_b(\bx)&&=
\hat{H}_{ab}\left(p_\mu+\frac{e}{c}A_\mu(\bx)\right)\psi_b(\bx)\nonumber\\
&&
+\int \rmd\br\ V^{\text{HF}}_{ab}(\bx,\br)\psi_b(\br),
\end{eqnarray}
where 
\begin{eqnarray}
&&V^{\text{HF}}_{ab}(\bx,\br)=
-V_{ad,cb}(\bx-\br)P_{dc}(\bx,\br)+\nonumber\\
&&\delta(\bx-\br)
\left(U_{ab}(\br)+
\int d^2\by V_{ab,cd}(\bx-\by)P_{dc}(\by,\by)\right) \ . \label{Eqn:VHF}
\end{eqnarray}
At self-consistency we can write the density matrix using the spectral representation as
\begin{eqnarray}\label{eq:P}
P_{ab}(\br,\br')=\oint_{\mathcal{C}}  \frac{\rmd z}{2\pi i} G_{ab}(\br,\br';z) \ ,  \label{Eqn:PfromG}
\end{eqnarray}
where we defined the Green's function
\begin{eqnarray}
G_{ab}(\br,\br';z)=\left\langle \br \right|\left(z-\hat{\mathcal{H}}^{\text{HF}}\right)_{ab}^{-1} \left| \br'\right\rangle \ . \label{Eqn:GInvH}
\end{eqnarray}
The contour of integration, $\mathcal{C}$, in the Eq.(\ref{eq:P}) is chosen to encircle a segment of the real axis with $E \leq \mu$. 

We proceed by assuming that we have found a self-consistent solution to the HF equations at $B\neq 0$. Our goal is then to find the series expansion of the grand potential $\Omega$, which is $E - \mu N$ at temperature $T = 0$, subject to the constraint of fixed chemical potential $\mu$. The coefficient of the $B$-linear term then determines the orbital magnetization and the coefficient of the $B^2$ term determines the orbital susceptibility.  In order to proceed, we find it convenient to isolate the phase factor associated with the (straight) line integral of the vector potential $\Phi(\br,\br')=\frac{e}{c\hbar}\int_\br^{\br'}\bA(\br'')\cdot d\br''$ where the charge of the electron is $-e$, and to express the Green's function as~\cite{PALeeOrbMag, raoux2015orbital, Gorkov61}
\begin{eqnarray}
G(\br,\br';z)=e^{i\Phi(\br,\br')}\widetilde{G}(\br,\br';z).
\end{eqnarray}
For concreteness, if we pick a Landau gauge, $\bA=Bx\hat{{\bf y}}$, then $\Phi(\br,\br')=\frac{eB}{2c\hbar}(r_x+r'_x)(r'_y-r_y)$. We stress that $\widetilde{G}$ carries non-trivial $B$ dependence and that we have {\it not} gauged away the $B$ field. 
The corresponding single particle density matrix is
$\widetilde{P}(\br,\br')=\oint_{\mathcal{C}}\frac{\rmd z}{2\pi i}
\widetilde{G}(\br,\br';z)$. The $B$ dependence of the total energy comes entirely from $\widetilde{P}$,
\begin{eqnarray}
&& \langle \mathcal{H}\rangle=\int \rmd \br \rmd \br' \left[ \delta(\br-\br') \left( \hat{H}_{ab}\left(p_\mu \right) +
U_{ab}(\br) \right) \widetilde{P}_{ba}(\br,\br') + \right. \nonumber\\
&& \left. \half V_{ab,cd}(\br-\br') \left(\widetilde{P}_{ba}(\br,\br)\widetilde{P}_{dc}(\br',\br')-\widetilde{P}_{da}(\br',\br)\widetilde{P}_{bc}(\br,\br')\right)\right]   \ .  \label{eq:EtotPtilde}
\end{eqnarray}
This is because $\hat{H}_{ab}\left(p_\mu+\frac{e}{c}A_\mu(\br)\right)e^{i\Phi(\br,\br')}=
e^{i\Phi(\br,\br')}\hat{H}_{ab}\left(p_\mu-\eps_{\mu\nu}\frac{eB}{2c}(r_\nu-r'_\nu)\right)$, where $\eps_{\mu\nu}$ is the antisymmetric Levi-Civita symbol, and because $\Lambda_{n,ab}^{\mu_1 \mu_2\ldots\mu_n }$ is symmetric under exchange of any two $\mu$-indices resulting in
$\eps_{\mu\nu}(r_\nu-r'_{\nu})$ effectively commuting with $\bp$ after the indices are summed. This allows us to move $\br-\br'$ to the left, next to the Dirac $\delta$ function, making such terms vanish.
The advantage of writing the total energy in terms of $\widetilde{P}$ is that $\left. \rmd \langle \mathcal{H} \rangle / \rmd B \right|_{B=0}$ can be expressed in terms of the action of the HF Hamiltonian $\hat{\mathcal{H}}^{\text{HF}}$ at $B=0$, which we denote by $\hat{\mathcal{H}}^{(0)\text{HF}}$,  on $\left. \rmd  \widetilde{P}/\rmd B\right|_{B=0}$.
The orbital magnetization $M$ can now be expressed as
\begin{eqnarray}
-M&& = \left. \frac{\rmd \Omega}{\rmd B} \right|_{B = 0} = \left.\frac{\rmd \langle \mathcal{H}\rangle}{\rmd B}\right|_{B=0}-
\mu \left.\frac{\rmd \langle\hat{N}\rangle}{\rmd B}\right|_{B=0}  \\
&& = \int \rmd\br \left\langle \br \right| \left(\hat{\mathcal{H}}^{(0)\text{HF}}_{ab} -\mu \delta_{ab} \right) \left. \frac{\rmd \hat{\widetilde{P}}_{ba}}{\rmd B}\right|_{B=0}\left|\br\right\rangle  \label{eq:mr} \\
\label{eq:MagdGdB}
&&=
\sum_{n}\oint_\mathcal{C}\frac{\rmd z}{2\pi i}  \left(E_n - \mu \right) \langle n | \left. \frac{\rmd \hat{\widetilde{G}}(z)}{\rmd B} \right|_{B=0} | n \rangle \ ,
\end{eqnarray}
where the sum is over the complete set of eigenstates, $|n\rangle$, of $\hat{\mathcal{H}}^{(0)\text{HF}}$, whose eigenvalues are $E_n$. 
The position space matrix elements of the Green's function operator $\hat{\widetilde{G}}(z)$ are $\widetilde{G}(\br,\br';z)$.
Next, we use the identity $\langle\br|\left(z-\hat{\mathcal{H}}^{\text{HF}}\right)\hat{G}(z)|\br'\rangle=\delta(\br-\br')$ which we multiply on both sides by $e^{i\Phi(\br',\br)}$. The right hand side remains $\delta(\br-\br')$ even after this multiplication and is therefore $B$ independent. Straightforward rearrangements then give
\begin{eqnarray}\label{eq:tildeDyson}
&&\int d\bx\left(z\delta(\br-\bx) \delta_{ab}  -\widetilde{V}^{\text{HF}}_{ab}(\br,\bx)\right)
e^{i\varphi(\br', \br, \bx)}
\widetilde{G}_{ba'}(\bx,\br';z)-\nonumber\\    
&&
\hat{H}_{ab}\left(p_\mu-\eps_{\mu\nu}\frac{eB}{2c}(r_\nu-r'_\nu)\right)\widetilde{G}_{ba'}(\br,\br';z)=\delta_{aa'}\delta(\br-\br'), \nonumber \\  \label{Eqn:Gtilde}
\end{eqnarray}
where 
\begin{eqnarray}
&&\widetilde{V}^{\text{HF}}_{ab}(\br, \bx) =-V_{ad,cb}(\br-\bx)\widetilde{P}_{dc}(\br,\bx)+\nonumber\\
&&\delta(\br-\bx)\left(U_{ab}(\br)+\int \rmd^2\by V_{ab,cd}(\br-\by) \widetilde{P}_{dc}(\by,\by)\right)
\end{eqnarray} and $e^{i\varphi(\br',\br,\bx)}=e^{i\Phi(\br',\br)}e^{i\Phi(\br,\bx)}
e^{i\Phi(\bx,\br')}
$.
We now wish to take the derivative of Eq.(\ref{eq:tildeDyson}) with respect to $B$ and then set $B=0$.
The gauge invariant phase factor  in the Eq.(\ref{eq:tildeDyson}) is determined by the magnetic flux through the triangle defined by the points $\br'$, $\br$ and $\bx$, and 
$e^{i\varphi(\br',\br,\bx)}=\exp\left(i\frac{eB}{2\hbar c}\hat{\bz}\cdot\left(\bx-\br\right)\times\left(\br'-\bx\right)\right)$. Its derivative at $B=0$ thus gives $i\frac{e}{2\hbar c}\eps_{\mu\nu}(x_\mu-r_\mu)(r'_\nu-x_\nu)$. The derivative of the single particle term $\hat{H}_{ab}$ at $B=0$ can be expressed as
$-\frac{i}{\hbar}\left[\hat{H}_{ab}\left(p_\mu\right),r_\mu\right]\eps_{\mu\nu}\frac{e}{2c}(r_\nu-r'_\nu)$.
We can rewrite expressions thus obtained in terms of operator matrix elements. For example  $(r'_\nu-x_\nu)\widetilde{G}_{ba'}(\bx,\br';z)=-\langle \bx|\left[\hat{x}_\nu,\hat{\widetilde{G}}_{ba'}(z)\right] |\br'\rangle$ and
$\widetilde{V}^{\text{HF}}_{ab}(\br,\bx)(x_\mu-r_\mu)=\langle \br| \left[\hat{\widetilde{V}}^{\text{HF}}_{ab},\hat{x}_{\mu}\right]|\bx\rangle$, where $\hat{x}_\mu$ is the position operator.
Moreover,
$\eps_{\mu\nu}\left[\hat{H}_{ab}\left(p_\mu\right),r_\mu\right](r_\nu-r'_\nu)
\widetilde{G}_{ba'}(\br,\br';z)=
\eps_{\mu\nu}\int d\bx \langle \br|\left[\hat{H}_{ab}\left(\hat{p}_\mu\right),\hat{x}_\mu\right] |\bx\rangle 
\langle \bx |\left[\hat{x}_{\nu},\hat{\widetilde{G}}_{ba'}(z)\right] | \br'\rangle$.
Finally, using $\left[\hat{G}^{(0)}(z),\hat{x}\right]=\hat{G}^{(0)}(z)
\left[\hat{\mathcal{H}}^{(0)\text{HF}},\hat{x}\right]
\hat{G}^{(0)}(z)
$, where at $B=0$ the Green's function operator $\hat{\widetilde{G}}$ is equal to $\hat{G}^{(0)}=\left(z-\hat{\mathcal{H}}^{(0)\text{HF}}\right)^{-1}$, we can use the derivative of the Eq.(\ref{eq:tildeDyson}) at $B=0$
to obtain an operator identity (see also Ref.~\cite{Tremblay14})
\begin{eqnarray}\label{eq:dGtildedB}
&&\left.\frac{d\hat{\widetilde{G}}_{b'a'}(z)}{dB}\right|_{B=0}=\hat{G}^{(0)}_{b'a}(z)\left.\frac{d\hat{\widetilde{V}}^{\text{HF}}_{ab}}{dB}\right|_{B=0}
\hat{G}^{(0)}_{ba'}(z)+\nonumber\\
&&\frac{ie}{2\hbar c}\eps_{\mu\nu}
\hat{G}^{(0)}_{b'a}(z)\left[\hat{\mathcal{H}}_{ab}^{(0)\text{HF}},\hat{x}_\mu\right]
\hat{G}_{bc}^{(0)}(z)\left[\hat{\mathcal{H}}^{(0)\text{HF}}_{cc'},\hat{x}_\nu\right]\hat{G}_{c'a'}^{(0)}(z). \nonumber \\   \label{Eqn:dGTildedB}
\end{eqnarray}
The operator identity can be used to substitute into Eq.(\ref{eq:MagdGdB}) in order to obtain an expression for the orbital magnetization $M$. 

We note that the Eq.(\ref{eq:dGtildedB}) is an integral equation which contains the (unknown) derivative of $\hat{\widetilde{G}}(z)$ implicitly in $\hat{\widetilde{V}}^{\text{HF}}$ because
\begin{eqnarray}
\left\langle \br\right| \left.\frac{d\hat{\widetilde{V}}^{\text{HF}}_{ab}}{dB}\right|_{B=0} \!\!\!\!\!\!\!\!\left|\bx\right\rangle&&\equiv
\delta(\br-\bx)\int d\by V_{ab,cd}(\br-\by)\left.\frac{d\widetilde{P}_{dc}(\by,\by)}{dB}\right|_{B=0}\nonumber\\
&&-V_{ad,cb}(\br-\bx)\left.\frac{d\widetilde{P}_{dc}(\br,\bx)}{dB}\right|_{B=0},
\end{eqnarray}
and because $\widetilde{P}$ is obtained from $\widetilde{G}$ by the contour integration. 
Fortunately, it is not necessary to solve the Eq.(\ref{eq:dGtildedB}) in order to obtain $M$ at zero temperature. 
That is because $\oint_{\mathcal{C}}\frac{dz}{2\pi i}\left(E_n-\mu\right)\langle n| 
\hat{G}^{(0)}(z)\left.\frac{d\hat{\widetilde{V}}^{\text{HF}}}{dB} \right|_{B=0}
\hat{G}^{(0)}(z)|n\rangle=0$ due to the coinciding poles of $\hat{G}^{(0)}(z)$ and the factor of  $E_n-\mu$ which eliminates any contribution from the Fermi surface~\cite{SM}.
The commutator of the HF Hamiltonian and the position operaror in the Eq.(\ref{eq:dGtildedB})
is related to the HF group velocity operator, which is Hermitian, and reads
\begin{eqnarray}
\hat{v}^{\text{HF}}_{\mu}=\frac{i}{\hbar}\left[\hat{\mathcal{H}}^{(0)\text{HF}},\hat{x}_\mu\right].
\end{eqnarray}
We can now write the orbital magnetization as $M=$
\begin{eqnarray}
&&\frac{ie\hbar}{2c}\eps_{\alpha\beta}
\sum_{n n'}\oint_\mathcal{C}\frac{dz}{2\pi i}  \left(E_n-\mu\right)
\frac{\langle n|\hat{v}^{\text{HF}}_{\alpha}|n'\rangle
\langle n'|\hat{v}^{\text{HF}}_{\beta}|n\rangle}{(z-E_n)^2(z-E_{n'})}=
\frac{ie\hbar}{2c}\times
\nonumber\\
&&\eps_{\alpha\beta}
\sum_{n\neq n'} \left(E_n-\mu\right)
\langle n|\hat{v}^{\text{HF}}_{\alpha}|n'\rangle
\langle n'|\hat{v}^{\text{HF}}_{\beta}|n\rangle\frac{n_F(E_{n'})-n_F(E_{n})}{(E_{n'}-E_n)^2},
\label{eq:M_in_n_basis}\nonumber\\
\end{eqnarray}
where $n_F(E_n)=1/\left(e^{(E_n-\mu)/T}+1\right)$ and $T$ is taken to $0$. The antisymmetry of the $\eps_{\alpha\beta}$, and the symmetry of the product of two diagonal matrix elements of the velocity operator, is the reason why the $n=n'$ term is omitted. 

If $\hat{\mathcal{H}}^{(0)\text{HF}}$ is periodic then its eigenstates  are Bloch waves, labeled by the crystal momentum $\bk$ residing within the first Brillouin zone, and a band index $n$,
\begin{eqnarray}
&&\langle \br|n,\bk\rangle =e^{i\bk\cdot\br}u_{n,\bk}(\br).
\end{eqnarray}
We made the dependence on the $\bk$ quantum number explicit and $u_{n,\bk}(\br)$ is the periodic part of the $n^{\text{th}}$ band Bloch function.
The velocity operator $\hat{v}^{\text{HF}}_{\alpha}$ is invariant under lattice translations and therefore it cannot mix different $\bk$'s. Moreover, we can write its matrix elements as
$ \left\langle n,\bk\right|\hat{v}^{\text{HF}}_{\mu}\left|n',\bk'\right\rangle=
\delta_{\bk\bk'}\frac{1}{\hbar}\left(
E_{n'}(\bk)-E_{n}(\bk)\right)\left\langle u_{n,\bk}\bigg|\frac{\partial u_{n',\bk}}{\partial k_\mu}\right\rangle$~\cite{SM}.
Substituting into the second line of the Eq.~(\ref{eq:M_in_n_basis}) we have
\begin{eqnarray}
    &&M = \frac{e}{2i\hbar c}\eps_{\alpha\beta} \sum_{\bk}\sum_{n\neq n'} ( E_n(\bk) - \mu ) \left\langle u_{n,\bk}\bigg|\frac{\partial u_{n',\bk}}{\partial k_\alpha}\right\rangle \times\nonumber\\  
    &&\left\langle u_{n',\bk}\bigg|\frac{\partial u_{n,\bk}}{\partial k_\beta}\right\rangle \left(n_F(E_{n'}(\bk))-n_F(E_{n}(\bk))\right)   \ . \label{Eqn:MFormula}
\end{eqnarray}
Both factors $\epsilon_{\alpha\beta} \left\langle u_{n,\bk}\bigg|\frac{\partial u_{n',\bk}}{\partial k_\alpha}\right\rangle \left\langle u_{n',\bk}\bigg|\frac{\partial u_{n,\bk}}{\partial k_\beta}\right\rangle$ and  $n_F(E_{n'}(\bk))-n_F(E_{n}(\bk))$ change sign under the interchange of $n$ and $n'$. Thus, the factor $E_n(\bk) - \mu$ can be symmetrized and replaced by $\frac12(E_n(\bk) + E_{n'}(\bk) - 2 \mu)$. After this replacement and noticing that the above expression is real due to the antisymmetric tensor $\epsilon_{\alpha\beta}$, we obtain the zero temperature limit of the Eq.~(12) in the Ref.~\cite{shi2007quantum}. The remaining steps follow the derivation in Ref.~\cite{shi2007quantum, SM}. Consequently, the expression for $M$ has the identical form as the Eq.~(13) in Ref.~\cite{shi2007quantum} as $T\rightarrow 0$, except the wavefunctions are the eigenfunctions of the self-consistent HF Hamiltonian at $B=0$. Thus, for $\hat{\mathcal{H}}^{(0)\text{HF}}(\bk)=e^{-i\bk\cdot\br}\hat{\mathcal{H}}^{(0)\text{HF}}e^{i\bk\cdot\br}$, we find
\begin{widetext}
\begin{align}
    M & = \frac{e}{2i\hbar c}\eps_{\alpha\beta} \sum_{\bk}\sum_{n} \left\langle \frac{\partial u_{n,\bk}}{\partial k_\alpha}\left|\hat{\mathcal{H}}^{(0)\text{HF}}(\bk)+E_n(\bk)-2\mu\right|\frac{\partial u_{n,\bk}}{\partial k_\beta}\right\rangle
n_F(E_{n}(\bk))  \  .   \label{eq:M_main_result}
\end{align}    
\end{widetext}

We emphasize that the derivation relies on the HF self-consistency and the completeness of the resulting spectrum, as encoded in Eqs.~(\ref{Eqn:VHF})--(\ref{Eqn:GInvH}) and (\ref{eq:tildeDyson}). Consequently, even when the focus is on the contribution to the orbital magnetization coming from a limited subset of bands near the Fermi energy --termed the active bands in the moire systems--  progressively more remote bands should be included in the $B=0$ HF calculation until the convergence of the active bands' contribution to the magnetization is achieved. In addition, the method introduced here can be applied to study aperiodic systems. While the final formula for $M$ presented above is written for a periodic system, the Eq.~\ref{Eqn:dGTildedB} does not rely on periodicity. When combined with the Eq.~\ref{eq:mr}, it allows for application to spatially inhomogeneous systems, including with disorder or open boundaries.

The methodology which we developed here allows us to go beyond the linear term in $B$ as well as nonzero $T$. The calculations of the orbital susceptibility in the interacting case within the HF approximation are involved and subject of a future publication. Here we just state the final answer,
\begin{eqnarray}
    \chi = \left.\frac{\pd M(B,\mu)}{\pd B} \right|_{B = 0}  = \chi_{\text{band}} + \chi_{\text{int}} \ ,  \label{Eqn:Chiformula} 
\end{eqnarray}
where the derivative is taken at the fixed $\mu$ and
\begin{widetext}
\begin{align}\label{eq:chi_main_result}
    \chi_{\text{band}} & =  \frac{\hbar^2 e^2}{12 c^2} \sum_n \oint_{\mathcal{C'}} \frac{\rmd z}{2 \pi i}   n_F(z) \left\langle n \left|  \hat{G}^{(0)}(z) \left( \hat{\mathcal{M}}^{\text{HF}} \right)^{-1}_{xx} \hat{G}^{(0)}(z) \left( \hat{\mathcal{M}}^{\text{HF}} \right)^{-1}_{yy} - \left( \hat{G}^{(0)}(z)  \left( \hat{\mathcal{M}}^{\text{HF}} \right)^{-1}_{xy} \right)^2  \right. \right.  \nonumber \\
    & \quad \left. \left.   - 4 \left( \left(  \hat{G}^{(0)}(z)  \hat{v}^{\text{HF}}_x \right)^2   \left( \hat{G}^{(0)}(z)  \hat{v}^{\text{HF}}_y \right)^2  - \left( \hat{G}^{(0)}(z) \hat{v}^{\text{HF}}_x \hat{G}^{(0)}(z)  \hat{v}^{\text{HF}}_y \right)^2 \right) \right| n \right\rangle\ ,   \\
    \chi_{\text{int}} & = i \frac{e \hbar}{2 c} \epsilon_{\mu\nu} \sum_n \oint_{\mcC'} \frac{\rmd z}{2\pi i} n_F(z) \left\langle n \left|  \hat G^{(0)}(z) \hat v^{\text{HF}}_{\mu} \hat G^{(0)}(z) \hat v^{\text{HF}}_{\nu} \hat G^{(0)}(z)  \left. \frac{\rmd \hat{\widetilde{V}}^{\text{HF}}}{\rmd B} \right|_{B = 0} \right| n \right\rangle \ .  \label{Eqn:ChiInt}
\end{align}
\end{widetext}
Here the integration contour $\mcC'$ is chosen to enclose the entire real axis, and the inverse effective mass operator $\left( \hat{\mathcal{M}}^{\text{HF}} \right)^{-1}$ is defined as
\begin{align}
   \left( \hat{\mathcal{M}}^{\text{HF}} \right)^{-1}_{\mu\nu} = \left( \frac{i}{\hbar} \right)^2 \left[ \left[ \hat{\mathcal{H}}^{(0)\text{HF}}, \hat{x}_\mu \right],  \hat{x}_\nu \right] \ , 
\end{align}
with $\mu$, $\nu = x$ or $y$. Similar to $M$, $\chi_{\text{band}}$ can be obtained from the expression for the susceptibility derived for non-interacting systems~\cite{raoux2015orbital}, with the zero-field Green function $\hat{G}^{(0)}$, the velocity operator $\hat{v}_{\mu}$, and the inverse effective mass operator $\left( \mathcal{M}^{-1} \right)_{\mu\nu}$ replaced by their self-consistent HF counterparts. In contrast, $\chi_{\text{int}}$ has no analogue in the non-interacting theory~\cite{SM}.

In order to numerically test our results, we analyze an interacting Rashba-like continuum model which is sufficiently simple to allow for some analytic progress and at the same time it contains non-trivial physics we wish to examine. Its single particle Hamiltonian takes the form $\hat{H}_{ab}(\mbf{p}) = \begin{pmatrix}
m_0 + b_2 \mbf{p}^2 & b_1(p_x - i p_y) \\
b_1(p_x + i p_y) & -m_0 + b_2 \mbf{p}^2
\end{pmatrix}$ and $U_{ab}(\br) = 0$.
We view this model as a massive Dirac particle, whose bands are known to contain non-zero Berry curvature and orbital magnetization, with additional (positive) $\mbf{p}^2$ terms that turn the valence band dispersion upward, guaranteeing non-infinite particle number density at a finite Fermi energy.
We choose the interaction potential to be the contact interaction whose real space form is $V_{ab,cd}(\br-\br')= U\delta(\br - \br') \delta_{ab} \delta_{cd} (1 - \delta_{ac})$.  These terms fall within the general interacting Hamiltonian in Eq.(\ref{eq:generalH}). The model has both continuous translation symmetry and continuous rotation symmetry.  We choose to rescale all lengths by $\hbar b_2/b_1$ and all energies by $b_1^2/b_2$. The single particle part at $B=0$ then contains only one dimensionless parameter, $\Delta = m_0 b_2/b_1^2$. Strength of $B$ field is described by dimensionless parameter $\phi/\phi_0 = eB \hbar^2b_2^2/(hb_1^2)$, which is the fraction of the flux quantum through the area $\hbar^2b_2^2/b_1^2$.

\begin{figure}[t]
    \centering
    \includegraphics[width=\linewidth]{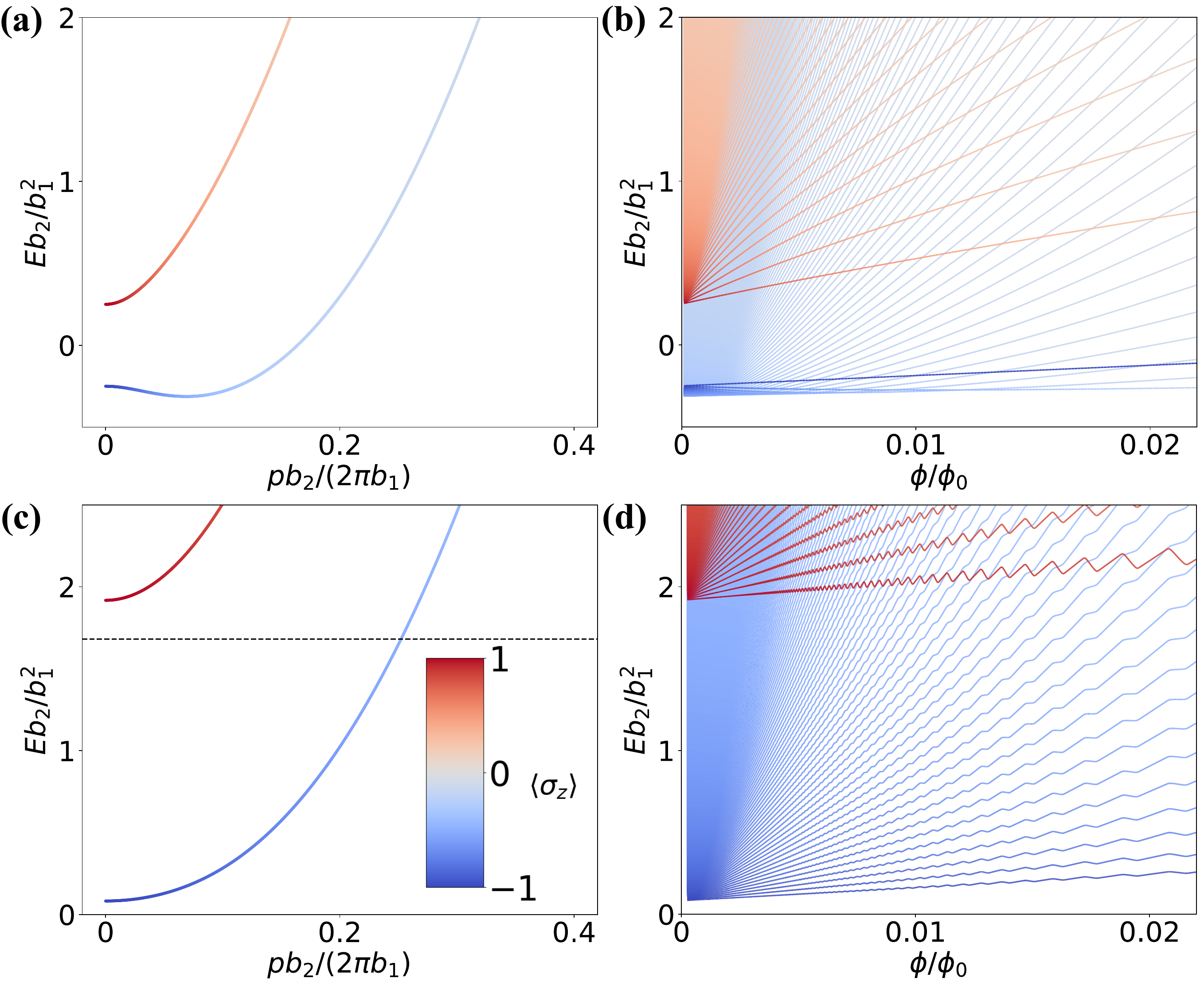}
    \caption{Energy spectrum of Rashba-like continuum model. Colors represent eigenstate polarization 
    $\langle \psi^{\dagger} \sigma_z \psi\rangle / \langle \psi^{\dagger} \psi\rangle$. (a) Non-interacting energy spectrum vs. momentum at $B=0$. (b) Non-interacting Landau level spectrum vs. $B$ field. (c) Finite temperature ($T$) HF energy spectrum at $B=0$ and fixed chemical potential $\mu$. The black dashed line marks $\mu$. (d) HF Landau level energy spectrum versus $B$ field at same $T$ and $\mu$. The oscillations within the spectrum are not numerical artifacts and come from the oscillatory solution to HF self-consistency equations. $T$ is chosen as low as $T = b_1^2/(380 k_B b_2)$. Parameters: $\{\Delta, U, \mu\} = \{1/4, 10 \hbar^2 b_2, 1.6816 b_1^2/b_2\}$. }
    \label{fig:intLLspectrum}
\end{figure}

\begin{figure}[t]
    \centering
    \includegraphics[width=0.9\linewidth]{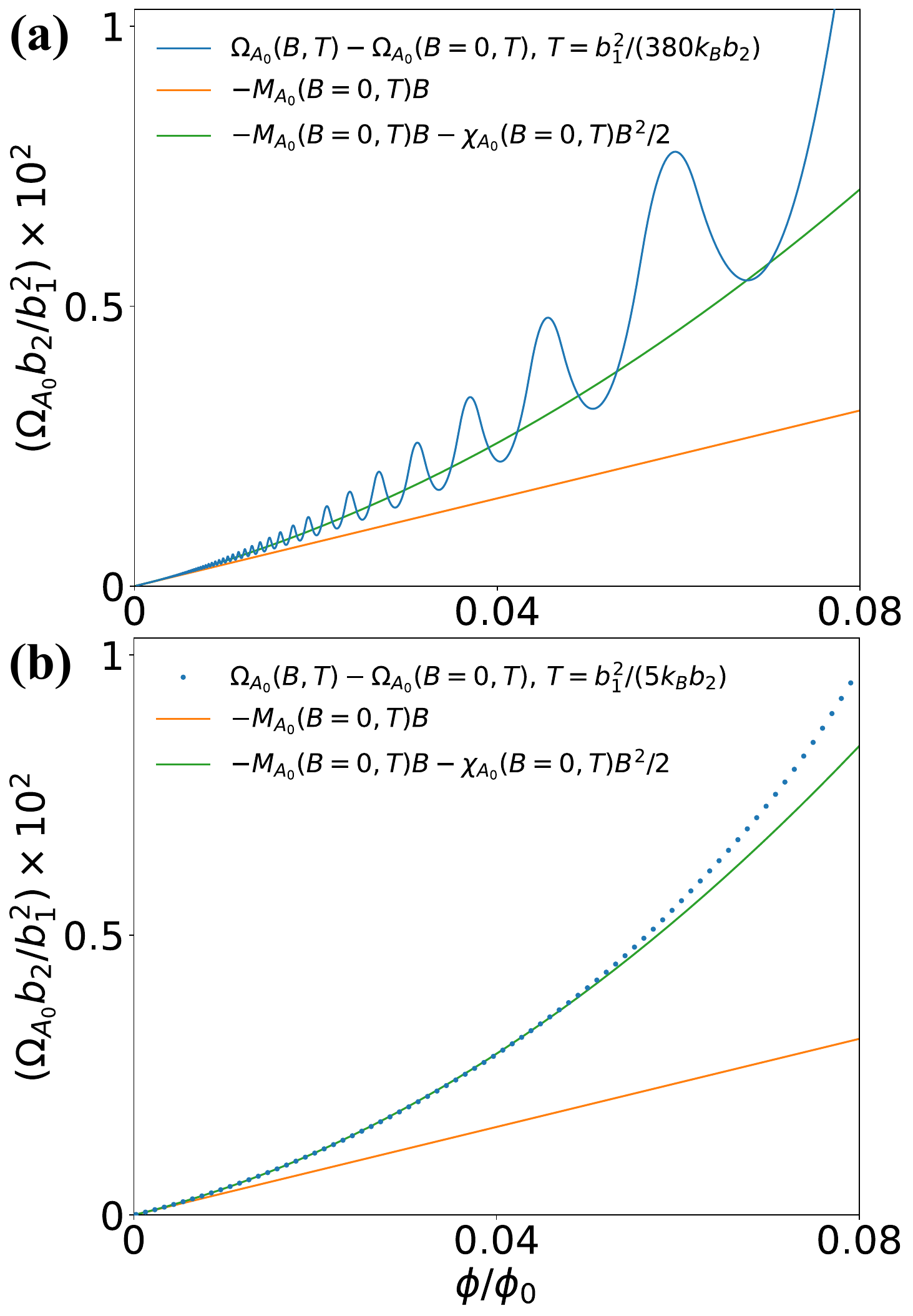}
    \caption{ HF grand potential $\Omega$ per unit area $A_0 = \hbar^2 b_2^2/b_1^2$ vs.~$B$ field, shown for (a) low $T$ and (b) intermediate $T$. In the small $B$ regime and for both temperatures, $\Omega$ approaches the linear extrapolation determined by 
    $M$ (Eq.~\ref{eq:M_main_result}) evaluated at $B = 0$. At larger magnetic fields, the inclusion of the quadratic correction associated with 
    $\chi$ (Eq.~(\ref{Eqn:Chiformula})--(\ref{Eqn:ChiInt})) at $B = 0$ further improves the agreement with $\Omega$, providing a non-trivial test of Eqs.~(\ref{eq:M_main_result})--(\ref{Eqn:ChiInt}). The parameters used are identical to those in Fig.~(\ref{fig:intLLspectrum}). The quantum oscillations visible at low $T$, manifested as the oscillatory behavior in $\Omega$, cannot be captured by the effective description based on zero-field $M$ and $\chi$ and are therefore beyond the scope of our present theory. }
    \label{fig:HFenergy}
\end{figure}

In the non-interacting case, we solved this model exactly at arbitrary $B$ using the Landau level basis, yielding both the energy spectrum and the associated wavefunctions. The spectra at $B=0$ and $B\neq0$ for $\Delta=1/4$ are shown in the Fig.~(\ref{fig:intLLspectrum}a) and (\ref{fig:intLLspectrum}b), respectively. We verified that $M$ and $\chi$ computed analytically directly at $B\neq0$ match the result obtained using the $B=0$ formulas in Eq.~(\ref{eq:M_main_result}) and Eq.~(\ref{eq:chi_main_result}) at small $B$; the details of this comparison will be presented in a future publication. Upon including the contact interaction, only the on-site density matrix $P_{ab}(\fvec r, \fvec r)$ enters the Hartree-Fock self-consistency equations at both $B = 0$ and $B \neq 0$. In the presence of the continuous (magnetic) translational symmetry $P_{ab}(\fvec r, \fvec r)$ reduces to constants independent of $\fvec r$. Furthermore, the rotational symmetry constrains the inter-flavor components $P_{a\neq b}(\fvec r, \fvec r)$ to vanish, since they would otherwise acquire a nontrivial phase under rotations. Consequently, the Hartree-Fock problem reduces to determining the two constant diagonal elements. The resulting interacting spectrum at $B= 0$ is shown in Fig.~\ref{fig:intLLspectrum}(c), with the chemical potential $\mu$ indicated by the dashed line. From the self-consistent solution at zero $B$ field, we compute the grand potential per area $\Omega_{A_0}(B = 0)$, the orbital magnetization $M_{A_0}(B = 0)$, and the orbital susceptibility $\chi_{A_0}(B = 0)$ using Eqs.~(\ref{eq:M_main_result}) and (\ref{eq:chi_main_result}). We then solve the Hartree-Fock equations at finite $B$ while keeping $\mu$ fixed at its zero-field value; the corresponding spectrum is shown in Fig.~\ref{fig:intLLspectrum}(d). Figure~\ref{fig:HFenergy} shows the grand-potential difference $\Omega_{A_0}(B) - \Omega_{A_0}(B = 0)$ as a function of $B$. The blue curve in Fig.~\ref{fig:HFenergy}a is at low $T$ thus displaying quantum oscillations and the blue dots in Fig.~\ref{fig:HFenergy}b are at higher $T$ where quantum oscillations are suppressed. These results are compared with the linear response prediction $-M_{A_0}(B = 0)B$ (orange solid line) and the quadratic expansion $-M_{A_0}(B = 0)B - \chi_{A_0}(B = 0) B^2 /2$ (green solid line). The excellent agreement in the $B\rightarrow 0$ limit provides a non-trivial validation of our main results.

In summary, in this work, we presented a method to compute the orbital magnetization $M$ and the orbital susceptibility $\chi$ for the interacting electrons within the self-consistent HF approximation. The resulting formulas involve only the self-consistent solution at $B=0$. The expressions for $M$ and $\chi$ in Eqs.~(\ref{eq:M_main_result})--(\ref{Eqn:ChiInt}) include additive contributions from all filled bands. In moire systems, restricting these sums to active bands only can, in general, omit significant contributions from remote bands~\footnote{We thank Y.~Zhang for sharing unpublished results (private communication).}. Applying these formulas to the study of interaction induced orbital magnets at small $B$ is significantly more efficient than performing the self-consistent HF approximation at $B\neq0$, providing a powerful framework for better understanding of this class of van der Waal heterostructures. 

\acknowledgments
We thank B. Andrei Bernevig, Di Xiao, Hui-Ke Jin, Wen-Yu He, Yi Zhang, Yuanyuan Zhao, Yves Kwan, Jonah Herzog-Arbeitman, and Ming-Rui Li for helpful discussions. M.W. acknowledges Taige Wang and Daniel Parker for useful discussions during the early development of codes for HF calculations at small $B$ field. J.~K.~acknowledges the support from the NSFC Grant No.~12074276, the NSFC Grant No.~12574159, the Double First-Class Initiative Fund of ShanghaiTech University, and the start-up grant of ShanghaiTech University. O.~V.~was funded in part by the Gordon and Betty Moore Foundation's EPiQS Initiative Grant GBMF11070 and is supported by a grant from the Simons Foundation SFI-MPS-NFS00006741-09. The simulations presented in this article were performed partly on computational resources managed and supported by Princeton University's Research Computing.

\bibliography{reNew}

\appendix
\begin{widetext}

	\newpage	
	
	\begin{center}
		\textbf{\large Supplemental Material for ``Orbital magnetization and magnetic susceptibility of interacting electrons''}
	\end{center}

\section{Velocity Operator}
In this section, we derive the matrix elements of the velocity operator in the Bloch state representation, i.e.
\begin{align}
    \langle n, \fvec k | \hat{v}^{\text{HF}}_{\mu} | n', \fvec k' \rangle \ .
\end{align}

Since the ground state at $B = 0$ is assumed to be translationally invariant, so is the Hartree-Fock Hamiltonian $\hat{\mcH}^{(0) \text{HF}}$. Therefore, the velocity operator, which is defined as the commutator between $\hat{\mcH}^{(0) \text{HF}}$ and the position operator, is also translationally invariant. As a consequence, 
\begin{align}
     \langle n, \fvec k | \hat{v}^{\text{HF}}_{\mu} | n', \fvec k' \rangle  \propto \delta_{\fvec k, \fvec k'} \ .
\end{align}

To derive the expression of the matrix elements of $\hat{v}_{\mu}^{\text{HF}}$, we use the operator $\hat{\mathcal{H}}^{(0) \text{HF}}(\fvec k)$, which is defined as 
\begin{align}
    \hat{\mathcal{H}}^{(0) \text{HF}}(\fvec k) = e^{- i \fvec k \cdot \hat{\fvec r}} \hat{\mathcal{H}}^{(0) \text{HF}} e^{i \fvec k \cdot \hat{\fvec r}}
\end{align}
where $\hat{\fvec r}$ is the position operator. Notice 
\begin{align}
   \hat{\mathcal{H}}^{(0) \text{HF}} | n, \fvec k \rangle & = E_n(\fvec k) | n, \fvec k \rangle =  e^{i \fvec k \cdot \hat{\fvec r}} E_n(\fvec k)  | u_{n, \fvec k} \rangle \nonumber \\
\mbox{Also}, \quad \hat{\mathcal{H}}^{(0) \text{HF}} | n, \fvec k \rangle   & = e^{i \fvec k \cdot \hat{\fvec r}} e^{- i \fvec k \cdot \hat{\fvec r}} \hat{\mathcal{H}}^{(0) \text{HF}} e^{i \fvec k \cdot \hat{\fvec r}} | u_{n, \fvec k} \rangle = e^{i \fvec k \cdot \hat{\fvec r}} \hat{\mathcal{H}}^{(0) \text{HF}}(\fvec k) | u_{n, \fvec k} \rangle \ .
\end{align}
Therefore, $| u_{n, \fvec k} \rangle$ is the eigenstate of $\hat{\mathcal{H}}^{(0) \text{HF}}(\fvec k)$ with the energy of $E_n(\fvec k)$.

Now, the matrix elements of the velocity operator are
\begin{align}
    & \quad \langle n, \fvec k | \hat{v}^{\text{HF}}_{\mu} | n', \fvec k \rangle  = \frac{i}{\hbar} \langle n, \fvec k | [ \hat{\mcH}^{(0) \text{HF}}, \hat{r}_{\mu} ] | n', \fvec k \rangle  = \frac{i}{\hbar} \langle u_{n, \fvec k} | e^{-i \fvec k \cdot \hat{\fvec r}} [ \hat{\mcH}^{(0) \text{HF}}, \hat{r}_{\mu} ] e^{i \fvec k \cdot \hat{\fvec r}} | u_{n', \fvec k} \rangle \nonumber \\
    & =  \frac1{\hbar} \left\langle u_{n, \fvec k} \left| \frac{\pd}{\pd k_{\mu}} \left( e^{-i \fvec k \cdot \hat{\fvec r}} \hat{\mcH}^{(0) \text{HF}} e^{i \fvec k \cdot \hat{\fvec r}} \right) \right| u_{n', \fvec k} \right\rangle = \frac1{\hbar} \left\langle u_{n, \fvec k} \left| \frac{\pd \hat{\mcH}^{(0) \text{HF}}(\fvec k)}{\pd k_{\mu}}   \right| u_{n', \fvec k} \right\rangle \nonumber \\
    & = \frac1{\hbar} \left(  \frac{\pd}{\pd k_{\mu}} \langle u_{n, \fvec k} | \hat{\mcH}^{(0) \text{HF}}(\fvec k) | u_{n', \fvec k} \rangle - \left\langle \frac{\pd u_{n, \fvec k}}{\pd k_{\mu}} \left|  \hat{\mcH}^{(0) \text{HF}}(\fvec k) \right| u_{n', \fvec k} \right\rangle - \left\langle u_{n, \fvec k} \left| \hat{\mcH}^{(0) \text{HF}}(\fvec k) \right| \frac{\pd u_{n', \fvec k}}{\pd k_{\mu}} \right\rangle  \right) \nonumber \\
    & = \frac1{\hbar} \frac{\pd E_n(\fvec k)}{\pd k_{\mu}} \delta_{n n'} - \frac{E_{n'}(\fvec k)}{\hbar} \left\langle \left. \frac{u_{n, \fvec k}}{\pd k_{\mu}} \right| u_{n', \fvec k} \right\rangle - \frac{E_n(\fvec k)}{\hbar} \left\langle u_{n, \fvec k} \left|  \frac{u_{n', \fvec k}}{\pd k_{\mu}} \right. \right\rangle 
\end{align}
In addition, we found that
\beq  
 \left\langle \left. \frac{u_{n, \fvec k}}{\pd k_{\mu}} \right| u_{n', \fvec k} \right\rangle = \frac{\pd}{\pd k_{\mu}} \langle u_{n, \fvec k} | u_{n', \fvec k} \rangle -  \left\langle u_{n, \fvec k} \left|  \frac{u_{n', \fvec k}}{\pd k_{\mu}} \right. \right\rangle  = -  \left\langle u_{n, \fvec k} \left|  \frac{u_{n', \fvec k}}{\pd k_{\mu}} \right. \right\rangle  \ .    \label{Eq:Formula1}
\eeq
Thus, the  matrix elements of the velocity operator can be written as
\begin{align}
    \langle n, \fvec k | \hat{v}^{\text{HF}}_{\mu} | n', \fvec k' \rangle  = \delta_{\fvec k, \fvec k'} \left( \frac1{\hbar} \frac{\pd E_n(\fvec k)}{\pd k_{\mu}} \delta_{n n'} + \frac{E_n(\fvec k) - E_{n'}(\fvec k)}{\hbar}  \left\langle u_{n, \fvec k} \left|  \frac{u_{n', \fvec k}}{\pd k_{\mu}} \right. \right\rangle   \right)  \  .
\end{align}

\section{Derivation of magnetization}
In this section, we present the necessary steps to derive the final expression of the orbital magnetization $M$. In the main text, we have already argued that 
\begin{align}
    M = \frac{e}{2i\hbar c}\eps_{\alpha\beta} \sum_{\bk}\sum_{n\neq n'} ( E_n(\bk) - \mu ) \left\langle u_{n,\bk}\bigg|\frac{\partial u_{n',\bk}}{\partial k_\alpha}\right\rangle     \left\langle u_{n',\bk}\bigg|\frac{\partial u_{n,\bk}}{\partial k_\beta}\right\rangle \left(n_F(E_{n'}(\bk))-n_F(E_{n}(\bk))\right) \ ,
\end{align}
and that the factor $E_n(\bk) - \mu$ in the above formula can be replaced by $\frac12(E_n(\bk) + E_{n'}(\bk) - 2 \mu)$. To obtain the final expression of $M$, notice that 
\begin{align}
    M & = \frac{e}{4i\hbar c}\eps_{\alpha\beta} \sum_{\bk}\sum_{n\neq n'} ( E_n(\bk) + E_{n'}(\bk) - 2\mu ) \left\langle u_{n,\bk}\bigg|\frac{\partial u_{n',\bk}}{\partial k_\alpha}\right\rangle     \left\langle u_{n',\bk}\bigg|\frac{\partial u_{n,\bk}}{\partial k_\beta}\right\rangle \left(n_F(E_{n'}(\bk))-n_F(E_{n}(\bk))\right)   \nonumber \\
    & =  \frac{e}{4i\hbar c} \eps_{\alpha\beta} \sum_{\bk}\sum_{n\neq n'} ( E_n(\bk) + E_{n'}(\bk) - 2\mu ) \left\langle u_{n,\bk}\bigg|\frac{\partial u_{n',\bk}}{\partial k_\alpha}\right\rangle     \left\langle u_{n',\bk}\bigg|\frac{\partial u_{n,\bk}}{\partial k_\beta}\right\rangle n_F(E_{n'}(\bk)) \nonumber \\
    & \quad -  \frac{e}{4i\hbar c} \eps_{\alpha\beta} \sum_{\bk}\sum_{n\neq n'} ( E_n(\bk) + E_{n'}(\bk) - 2\mu ) \left\langle u_{n,\bk}\bigg|\frac{\partial u_{n',\bk}}{\partial k_\alpha}\right\rangle     \left\langle u_{n',\bk}\bigg|\frac{\partial u_{n,\bk}}{\partial k_\beta}\right\rangle n_F(E_{n}(\bk))  \  .  \label{Eq:MStep1}
\end{align}
The first term in the above formula can be simplified as
\begin{align}
    & \quad \sum_{n \neq n'} ( E_n(\bk) + E_{n'}(\bk) - 2\mu ) \left\langle u_{n,\bk}\bigg|\frac{\partial u_{n',\bk}}{\partial k_\alpha} \right\rangle     \left\langle u_{n',\bk}\bigg|\frac{\partial u_{n,\bk}}{\partial k_\beta}\right\rangle n_F(E_{n'}(\bk)) \nonumber \\
    & = - \sum_{n,  n'} ( E_n(\bk) + E_{n'}(\bk) - 2\mu ) \left\langle u_{n,\bk}\bigg|\frac{\partial u_{n',\bk}}{\partial k_\alpha} \right\rangle     \left\langle \frac{\partial u_{n',\bk}}{\partial k_\beta} \bigg|  u_{n,\bk}  \right\rangle n_F(E_{n'}(\bk)) \nonumber \\
    & \quad - \sum_n 2(E_n(\fvec k) - \mu)  \left\langle u_{n,\bk}\bigg|\frac{\partial u_{n,\bk}}{\partial k_\alpha} \right\rangle   \left\langle u_{n,\bk}\bigg|\frac{\partial u_{n,\bk}}{\partial k_\beta}\right\rangle n_F(E_{n}(\bk)) \nonumber \\
    & = - \sum_{n'} \left\langle \frac{\partial u_{n',\bk}}{\partial k_\beta} \bigg| (  E_{n'}(\fvec k) + \hat{\mathcal{H}}^{(0) \text{HF}}(\fvec k) - 2 \mu ) \bigg| \frac{\partial u_{n',\bk}}{\partial k_\alpha}   \right\rangle n_F(E_{n'}(\fvec k)) \nonumber \\
    & \quad - \sum_n 2(E_n(\fvec k) - \mu)  \left\langle u_{n,\bk}\bigg|\frac{\partial u_{n,\bk}}{\partial k_\alpha} \right\rangle     \left\langle u_{n,\bk}\bigg|\frac{\partial u_{n,\bk}}{\partial k_\beta}\right\rangle n_F(E_{n}(\bk))   \  .  \label{Eq:MStep2}
\end{align}
where Eq.~(\ref{Eq:Formula1}) has been applied in the second step. The same approach can be applied to simplify the second term in Eq.~\ref{Eq:MStep1}, giving
\begin{align}
    & \quad \sum_{n\neq n'} ( E_n(\bk) + E_{n'}(\bk) - 2\mu ) \left\langle u_{n,\bk}\bigg|\frac{\partial u_{n',\bk}}{\partial k_\alpha}\right\rangle     \left\langle u_{n',\bk}\bigg|\frac{\partial u_{n,\bk}}{\partial k_\beta}\right\rangle n_F(E_{n}(\bk)) \nonumber \\
   & = - \sum_n  \left\langle \frac{\partial u_{n,\bk}}{\partial k_\alpha} \bigg| (  E_n(\fvec k) + \hat{\mathcal{H}}^{(0) \text{HF}}(\fvec k) - 2 \mu ) \bigg| \frac{\partial u_{n,\bk}}{\partial k_\beta}   \right\rangle n_F(E_n(\fvec k)) \nonumber \\
   & \quad - \sum_n 2(E_n(\fvec k) - \mu)  \left\langle u_{n,\bk}\bigg|\frac{\partial u_{n,\bk}}{\partial k_\alpha} \right\rangle     \left\langle u_{n,\bk}\bigg|\frac{\partial u_{n,\bk}}{\partial k_\beta}\right\rangle n_F(E_{n}(\bk)) \ .   \label{Eq:MStep3}
\end{align}
Substituting Eqs.~(\ref{Eq:MStep2}) and (\ref{Eq:MStep3}) into Eq.~(\ref{Eq:MStep1}), we obtain 
\begin{align}
        M & = \frac{e}{2i\hbar c}\eps_{\alpha\beta} \sum_{\bk}\sum_{n} \left\langle \frac{\partial u_{n,\bk}}{\partial k_\alpha}\left|\mathcal{H}^{(0)\text{HF}}(\bk)+E_n(\bk)-2\mu\right|\frac{\partial u_{n,\bk}}{\partial k_\beta}\right\rangle n_F(E_n(\fvec k)) \ ,
\end{align}
which is the final expression for the orbital magnetization presented in the main text.

\section{Contour integral}
In the main text, we have stated that the term
\begin{align}
    I = \oint_{\mathcal{C}}\frac{dz}{2\pi i} \left(E_n-\mu\right) \langle n|  \hat{G}^{(0)}(z)\left.\frac{\rmd \hat{\widetilde{V}}^{\text{HF}}}{\rmd B} \right|_{B=0} \hat{G}^{(0)}(z) | n \rangle \label{Eq:Integral}
\end{align}
vanishes due to the factor $E_n - \mu$. In this section, we provide the explicit calculation for this statement.

As mentioned in the text, the contour $\mcC$ is chosen to encircle a part of the real axis, so that all the Hartree-Fock energies below the Fermi surface are inside this contour. The integrated function in Eq.~\ref{Eq:Integral} contains an order-$2$ pole at $z = E_n$ because of the two Green functions. If the energy is below (or above) the Fermi surface, this pole is inside (or outside) the contour, and thus the contour integral vanishes simply by Cauchy's integral theorem. If the energy $E_n = \mu$ is at the Fermi surface, this pole is at the contour. For this case, we consider the Cauchy principal value of this integral, i.e.
\begin{align}
    I & = \left(E_n-\mu\right)   \lim_{\delta \rightarrow 0^+}  \left( \int_{- \infty - i \delta}^{\mu - i \delta} \frac{\rmd z}{2\pi i}  \ + \int_{\mu + i \delta}^{\infty + i\delta} \frac{\rmd z}{2\pi i} \  \right) \frac1{(z - E_n)^2} \langle n|  \left.\frac{\rmd \hat{\widetilde{V}}^{\text{HF}}}{\rmd B} \right|_{B=0} | n \rangle  \nonumber \\
    & = \frac{E_n - \mu}{2 \pi i} \lim_{\delta \rightarrow 0^+}  \left( - \frac1{\mu - i \delta - E_n} + \frac1{\mu + i \delta - E_n}  \right) \langle n|  \left.\frac{\rmd \hat{\widetilde{V}}^{\text{HF}}}{\rmd B} \right|_{B=0} | n \rangle \nonumber \\
    & =  (E_n - \mu) \lim_{\delta \rightarrow 0^+} \frac1{2 \pi i} \frac{2i \delta}{ (\mu - E_n)^2 + \delta^2  } \langle n|  \left.\frac{\rmd \hat{\widetilde{V}}^{\text{HF}}}{\rmd B} \right|_{B=0} | n \rangle = (E_n - \mu) \delta(E_n - \mu) \langle n|  \left.\frac{\rmd \hat{\widetilde{V}}^{\text{HF}}}{\rmd B} \right|_{B=0} | n \rangle = 0
\end{align}
Therefore, the contour integral in Eq.~\ref{Eq:Integral} vanishes and thus has no contribution to the orbital magnetization at $T = 0$.

\section{Orbital Susceptibility}
In the first version, we write $\chi = \chi_1 + \chi_2 + \chi_3$, where
\begin{align}\label{eq:chi1}
    \chi_1 & =  \frac{\hbar^2 e^2}{12 c^2} \sum_n \oint_{\mathcal{C}} \frac{\rmd z}{2 \pi i}   \langle n |  \hat{G}^{(0)}(z) \left( \hat{\mathcal{M}}^{\text{HF}} \right)^{-1}_{xx} \hat{G}^{(0)}(z) \left( \hat{\mathcal{M}}^{\text{HF}} \right)^{-1}_{yy} - \hat{G}^{(0)}(z)  \left( \hat{\mathcal{M}}^{\text{HF}} \right)^{-1}_{xy} \hat{G}^{(0)}(z)  \left( \hat{\mathcal{M}}^{\text{HF}} \right)^{-1}_{xy} \nonumber \\
    & \quad   - 4 \left( \hat{G}^{(0)}(z)  \hat{v}^{\text{HF}}_x  \hat{G}^{(0)}(z)  \hat{v}^{\text{HF}}_x \hat{G}^{(0)}(z)  \hat{v}^{\text{HF}}_y \hat{G}^{(0)}(z)  \hat{v}^{\text{HF}}_y - \hat{G}^{(0)}(z) \hat{v}^{\text{HF}}_x \hat{G}^{(0)}(z)  \hat{v}^{\text{HF}}_y \hat{G}^{(0)}(z)  \hat{v}^{\text{HF}}_x \hat{G}^{(0)}(z)  \hat{v}^{\text{HF}}_y \right)  | n\rangle, \\
    \chi_2 & =  \sum_n \oint_{\mathcal{C}} \frac{\rmd z}{2 \pi i}   \left\langle n \left| \hat{G}^{(0)}(z) \left. \frac{\rmd \hat{\widetilde{V}}^{HF}}{\rmd B} \right|_{B = 0} \hat{G}^{(0)}(z)  \left. \frac{\rmd \hat{\widetilde{V}}^{\text{HF}}}{\rmd B} \right|_{B = 0} \right| n \right\rangle, \\
    \label{eq:chi3}
    \chi_3 & =  - \int \rmd \fvec r\ \rmd \fvec r'\ V_{ab, cd}(\fvec r - \fvec r') \left( \left. \frac{\rmd \widetilde{P}_{ba}(\fvec r, \fvec r)}{\rmd B} \right|_{B = 0} \left. \frac{\rmd \widetilde{P}_{dc}(\fvec r', \fvec r')}{\rmd B} \right|_{B = 0} - \left. \frac{\rmd \widetilde{P}_{da}(\fvec r', \fvec r)}{\rmd B} \right|_{B = 0} \left. \frac{\rmd \widetilde{P}_{bc}(\fvec r, \fvec r')}{\rmd B}  \right|_{B = 0}   \right) \ ,  
\end{align}
In the revised manuscript, we decompose the orbital magnetic susceptibility into $\chi_{\text{band}}$ and $\chi_{\text{int}}$. As defined in the main text, $\chi_{\text{band}} = \chi_1$, which can be obtained from the corresponding non-interacting expression by replacing the zero-field Green function, velocity operators, and inverse mass operator with their self-consistent HF counterparts. By contrast, $\chi_{\text{int}}$ depends explicitly on the electron-electron interactions and thus has no analogue in the non-interacting theory. We will demonstrate in a subsequent work that $\chi_{\text{int}} = \chi_2 + \chi_3$.

\end{widetext}

\end{document}